\documentclass[preprint]{aastex}
\usepackage{epstopdf}
\usepackage{amsmath,amssymb}
\usepackage{graphicx,dblfloatfix}

\begin{document}

\title{HD 106906:  A Case Study for External Perturbations of a Debris Disk}

\author{Erika R. Nesvold\altaffilmark{1},
Smadar Naoz\altaffilmark{2,3},
Michael Fitzgerald\altaffilmark{2}}
\altaffiltext{1}{Department of Terrestrial Magnetism, Carnegie Institution for Science, 5241 Broad Branch Rd, Washington, DC 20015}
\altaffiltext{2}{Department of Physics and Astronomy, UCLA, 475 Portola Plaza, Los Angeles, CA 90095}
\altaffiltext{3}{Mani L. Bhaumik Institute for Theoretical Physics, Department of Physics and Astronomy, UCLA, Los Angeles, CA 90095}

\begin{abstract}
Models of debris disk morphology are often focused on the effects of a planet orbiting interior to or within the disk. Nonetheless, an exterior planetary-mass perturber can also excite eccentricities in a debris disk, via Laplace-Lagrange secular perturbations in the coplanar case or Kozai-Lidov perturbations for mutually inclined companions and disks. HD 106906 is an ideal example of such a a system, as it harbors a confirmed exterior 11 $\rm M_{\rm Jup}$ companion at a projected separation of 650 au outside a resolved, asymmetric disk. We use collisional and dynamical simulations to investigate the interactions between the disk and the companion, and to use the disk's observed morphology to place constraints on the companion's orbit. We conclude that the disk's observed morphology is consistent with perturbations from the observed exterior companion. Generalizing this result, we suggest that exterior perturbers, as well as interior planets, should be considered when investigating the cause of observed asymmetries in a debris disk.
\end{abstract}

\section{Introduction}
\label{sec:introduction}

Circumstellar debris disks are produced by the rocky and icy material leftover from the formation of the star and any planets in the system. To date, over 1700 debris disks have been detected via the excess infrared emission in their star's spectral energy distribution \citep{Cotten2016}, and over 40 have been resolved with optical or infrared imaging (http://circumstellardisks.org). The architecture of the underlying planetary system can leave a distinct imprint on the morphology of a debris disk \citep{Mouillet1997,Wyatt1999a,Matthews2014,Nesvold2015,Lee2016,Nesvold2016}. 

Modeling debris disk morphology is often focused on the effects of a planetary-mass perturber orbiting interior to or within the disk \citep{Mouillet1997,Chiang2009,Pearce2015,Nesvold2015a}. Nonetheless, exterior companions have been detected and inferred for several systems \citep{Rodriguez2012,Bailey2014,Mawet2015}, and dynamical modeling suggests that an exterior perturber can also excite eccentricities of the particles in a debris disk, via Laplace-Lagrange secular perturbations in the near-coplanar case \citep{Thebault2012a} or Kozai-Lidov perturbations for mutually inclined companions and disks \citep{Nesvold2016}, inducing asymmetries in the disk and triggering a collisional cascade. Collisions between the disk particles produce smaller dust grains whose thermal emission or scattered light can then be spatially resolved with infrared or optical imaging \citep{Wyatt2008}.

HD 106906 is an ideal example of a system with an exterior perturber, as it harbors a confirmed exterior companion with a model-atmosphere-derived mass of 11 $\rm M_{\rm Jup}$ at a projected separation of $650$ AU outside a resolved disk \citep{Bailey2014}. Scattered-light imaging of the disk with the Gemini Planet Imager (GPI), the Hubble Space Telescope's Advanced Camera for Surverys (HST/ACS), and SPHERE has revealed that the disk is a ring viewed nearly edge-on (inclination $\sim85^{\circ}$), with a inner region cleared of small dust grains \citep{Kalas2015, Lagrange2016}. These observations noted four major features of the disk morphology and system geometry: 
\begin{enumerate}
\item The position angle (PA) of the disk is oriented $\sim21^{\circ}$ counterclockwise from the position angle of the companion, which constrains the orbit of the companion relative to the disk. 
\item The inner disk has little to no vertical extension. While \citet{Kalas2015} tentatively suggested the presence of a ``warp'' in the disk's vertical structure on the west side of the disk, this warp was not confirmed by \citet{Lagrange2016}. This lack of vertical extension indicates that the inclinations of the disk particles have not been excited. 
\item The east side of the disk is brighter than the west side in GPI and SPHERE near-infrared images. 
\item \citet{Kalas2015} observed a faint extension on the west side of the disk out to nearly 500 au, but only diffuse emission on the east side . 
\end{enumerate}
These latter two features indicate that the disk may be an eccentric ring, which will exhibit a brightness asymmetry towards the pericenter side \citep[``pericenter glow'',][]{Wyatt1999a} and a faint, extended tail towards the apocenter side \citep{Lee2016}.

We modeled the HD 106906 system to demonstrate that the observed exterior companion can shape the disk into a flat, eccentric ring, and that all four of these morphological features can be reproduced without invoking the presence of a second companion. We also used the observed features and asymmetries of the HD 106906 disk to place constraints on the orbit of the observed companion. In Section \ref{sec:simulations}, we describe the collisional and dynamical simulations we performed of the parent bodies and dust grains in the HD 106906 disk. In Section \ref{sec:results}, we present the simulated brightness images produced by our simulations for comparison with observations. In \ref{sec:time}, we discuss the implication of these results and show how they can be used to constrain the orbit of HD 106906b. In Section \ref{sec:conclusions}, we summarize our conclusions and suggest opportunities for future work.

\section{Simulations}
\label{sec:simulations}

Given that collisions between particles in a disk with sufficiently high optical depth ($L_{\rm IR}/L_{*} \approx1.4\times10^{-3}$ for HD 106906 \citep{Chen2011}) will both produce the small grains seen in observations and may affect the dynamics of the disk, we simulated the HD 106906 system using the Superparticle-Method Algorithm for Collisions in Kuiper belts and debris disks \citep[SMACK,][]{Nesvold2013}. We then recorded the dust-producing encounters between parent bodies, simulated the orbits of the generated dust grains under the influence of radiative forces following the method of \citet{Lee2016}, then simulated the surface brightness of the dust using a Henyey-Greenstein scattering phase function \citep{Henyey1941}.

\subsection{SMACK Model}

SMACK is based on the $N$-body integrator REBOUND \citep{Rein2012}, but approximates each particle in the integrator as a collection of bodies with a range of sizes between 1 mm and 10 cm in diameter, traveling on the same orbit. This group of bodies is called a ``superparticle'' and is characterized by a size distribution, position, and velocity. The superparticles act as test particles in the integration, and orbit the star under the influence of perturbations by any planets in the simulation. Each superparticle is approximated as a sphere with some finite radius. When REBOUND detects that two superparticles are overlapping in space, SMACK statistically calculates the number of bodies within each superparticle that will collide and fragment, removes these bodies from their size distributions, and redistributes the fragments. SMACK also corrects the trajectories of the parent superparticles to conserve angular momentum and energy, compensating for the kinetic energy lost to fragmentation.

The parameters for the SMACK simulation of the HD 106906 system described in this work are listed in Table \ref{tab:init}. The masses of the star and companion were 2.5 $\rm{M}_{\odot}$ and 11 $\rm{M_{Jup}}$, respectively. The initial semi-major axis range of $65-85$ au for the 10,000 superparticles in the simulated disk was chosen in anticipation that the disk would spread during the 15 Myr course of the simulation. The radial extent of the HD 106906 ring as observed in scattered-light imaging is $\sim50-100$ au \citep{Kalas2015,Lagrange2016}. The orbital parameters of the companion were chosen such that the gravitational perturbations from the companion would excite the eccentricities of the disk particles without causing a vertical extension of the disk on the timescale of the system's age, and such that the simulated companion's orbit could reproduce the position of the observed companion.

\begin{table}
\caption{Initial conditions of the disk and companion for the simulation. \label{tab:init}}
\begin{tabular}{lcc} 
Parameter	 & Initial Disk Values & HD 106906b \\
\hline
Semi-Major Axis (au) 			& $65-85$			 & 700  \\
Eccentricity 					& $0-0.01$ 		 & 0.7  \\
Inclination ($^{\circ}$) 			& $0-0.29$		 & 8.5 \\
Longitude of Nodes ($^{\circ}$)		& $0-360$			 & 90 \\
Argument of Pericenter ($^{\circ}$) 	& $0-360$			 & $-90$ \\
Optical depth					& $1.4\times10^{-3}$ & --	\\
Density (g~cm$^{-3}$) 			& 1.0	 			 & -- 	\\
\end{tabular}
\end{table}

\subsection{SMACK vs. Collisionless $N$-Body}

\begin{figure}
\includegraphics[width=\columnwidth]{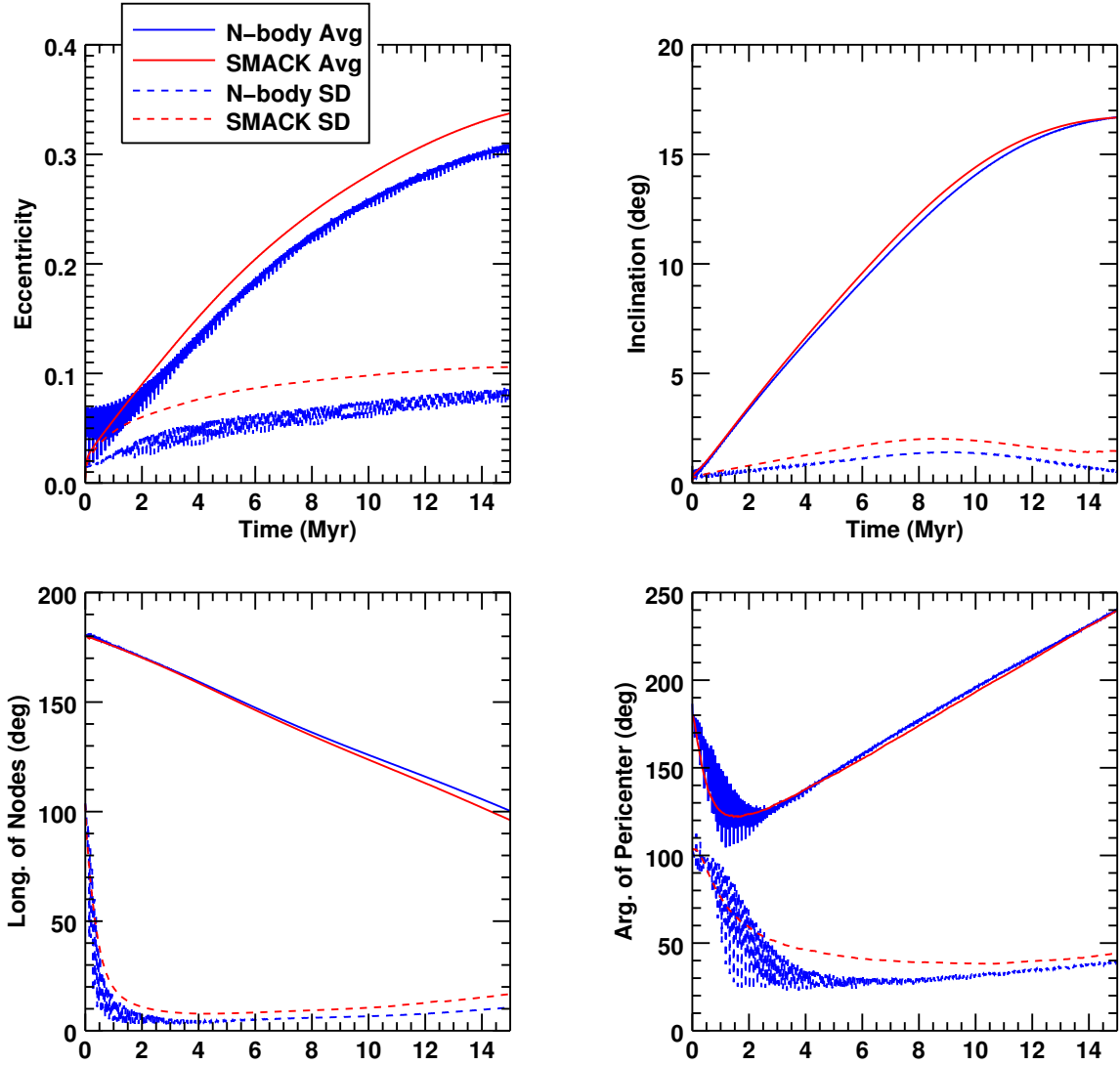}
\caption{\label{fig:orbelcompare} Time evolution of the average eccentricity, inclination, longitude of nodes, and argument of pericenter for disk particles in the SMACK and collisionless $N$-body simulations described in this section. The dashed lines indicate the standard deviation of each orbital element for each simulation. The variation between the two simulations is $\leq10\%$ for each orbital element's average}.
\end{figure}

To measure the effects of collisions on the dynamics of the disk particles, we also performed a collisionless $N$-body simulation of the disk using the Wisdom-Holman integrator of REBOUND with collision detection and resolution turned off. The collisionless $N$-body simulation used the same companion and disk parameters as the SMACK simulation (Table \ref{tab:init}), with 10,000 test particles to represent the disk. Figure \ref{fig:orbelcompare} shows the time evolution of the simulated disk's average eccentricity, inclination, longitude of nodes, and argument of pericenter. Although there are small variations, most notably in the average eccentricity, between the SMACK simulation and the collisionless $N$-body simulation, the maximum difference for each parameter is $\leq10\%$, indicating that fragmenting collisions have a minimal effect on the dynamics of this system.

\subsection{Dust Model}
\label{sec:dust}

To generate the simulated images of the dust grains in the HD 106906 system, we adapted the method of \citet{Lee2016}, which extended the dust orbit calculations of \citet{Wyatt1999a} to include estimates of the surface brightness. Our SMACK simulation output the locations of dust-producing collisions during the 15 Myr simulation, as well as the orbits of the parent bodies producing the dust. We selected the first $10^4$ dust production events occurring after time $t=5$ Myr. For each dust production event, we generated 10 dust orbits, each with a $\beta$ value randomly chosen from a power-law distribution with index 3/2 (where $\beta\approx F_{rad}/F_{grav}$ represents the ratio of the radiative and gravitational forces acting on a dust grain). The maximum possible value for the $\beta$ value was set by the parent body's orbit:
\begin{equation} \beta_{\rm max} = \frac{1-e_{\rm p}}{2(1+e_{\rm p} \cos f_{\rm p})}, \end{equation}
where $e_{\rm p}$ and $f_{\rm p}$ are the parent body's eccentricity and true anomaly, respectively. The semi-major axis $a$, eccentricity $e$, and argument of pericenter $\omega$ of each dust orbit are given by the parent body's orbit (specifically $a_{\rm p}$, $e_{\rm p}$, and $f_{\rm p}$) and the $\beta$ value assigned to the orbit:
\begin{equation} a = \frac{a_{\rm p}(1-e_{\rm p}^2)(1-\beta)}{1-e_{\rm p}^2-2\beta(1-e_{\rm p}\cos f_{\rm p})}, \end{equation}
\begin{equation} e = \frac{\sqrt{e_{\rm p}^2+2\beta e_{\rm p} \cos f_{\rm p} + \beta^2}}{1-\beta}, \end{equation}
\begin{equation} \omega = \omega_{\rm p} + \arctan\left(\frac{\beta \sin f_{\rm p}}{e_{\rm p}+\beta \cos f_{\rm p}}\right). \end{equation}
The inclination $i$ and longitude of nodes $\Omega$ of the dust orbit was set to be equal to the corresponding values for the parent body, $i_{\rm p}$ and $\Omega_{\rm p}$, respectively. For each dust orbit, we generated 10 dust grains, and assigned each a mean anomaly selected randomly from a uniform distribution between 0 and $360^{\circ}$. Thus, each dust production event produced 100 final dust grain locations.

After constructing the dust population from the SMACK results, we simulated the surface brightness of the dust using $\phi(g,\theta)/\beta^2 r^2$, where $\phi(g,\theta)$ is the Henyey-Greenstein scattering phase function with asymmetry parameter $g$, $\theta$ is the angle between the dust grain and the observer's line-of-sight (with the vertex at the star), and $r$ is the distance between the dust grain and the star. Following \citet{Lee2016}, we used $g=0.5$.

\section{Results}
\label{sec:results}

\begin{figure}
\includegraphics[width=\linewidth]{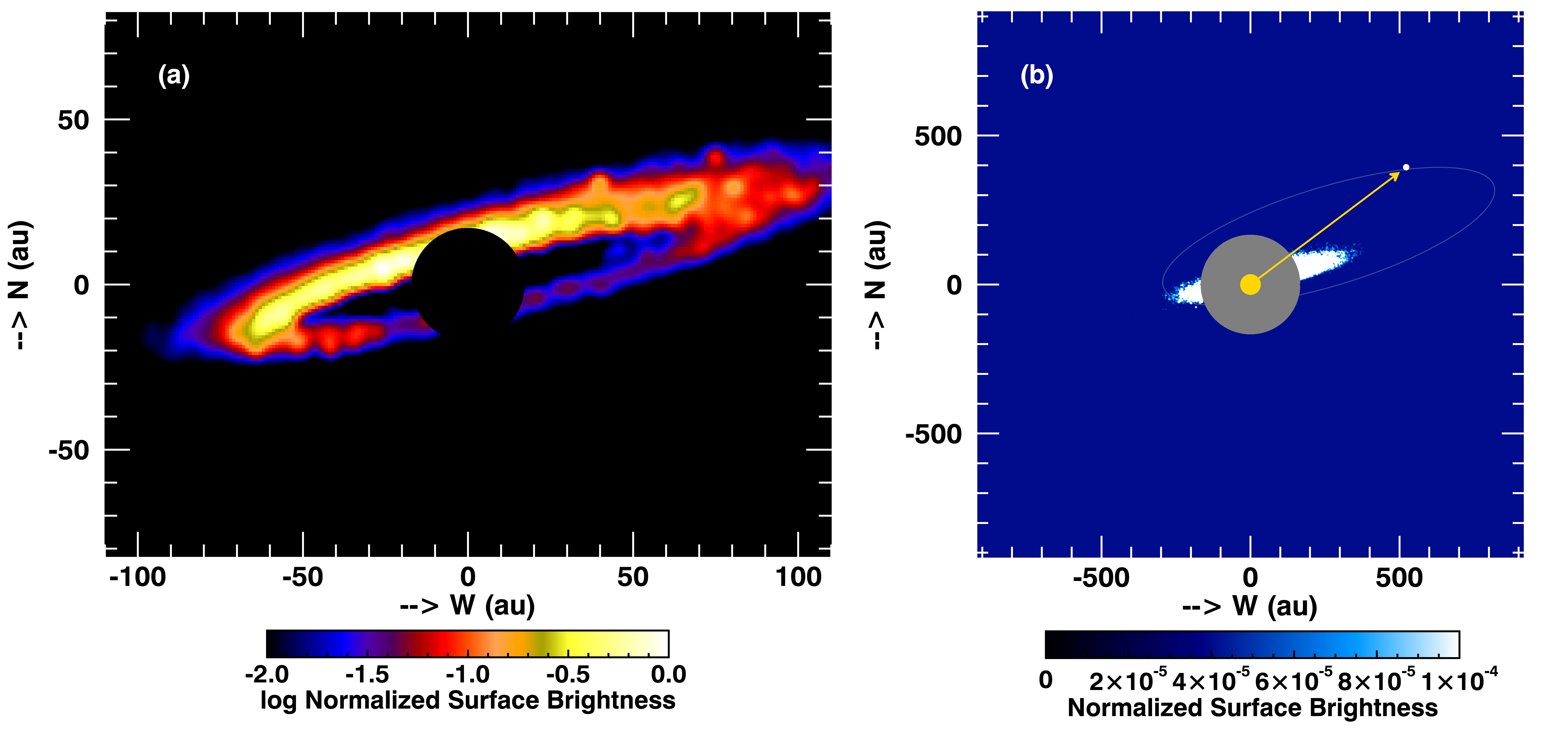}
\caption{\label{fig:combined} Simulated surface brightness of the SMACK-simulated dust ring after 5 Myr of perturbations from a companion at semi-major axis $a_{\rm pl}=700$ au, eccentricity $e_{\rm pl}=0.7$, and inclination $i_{\rm pl} = 8.5^{\circ}$. The viewing inclination is $\sim5^{\circ}$ from edge-on and the pericenter side of the disk is towards the east. (a) The field of view and simulated coronagraphic mask were chosen for comparison with Figure 1 of the \citet{Lagrange2016}. Pericenter glow causes the east side of the disk to appear brighter. (b) The field of view and simulated coronagraphic mask were chosen for comparison with Figure 3 of \citet{Kalas2015}. The grey line in (b) indicates the orbit of the simulated companion, while the white dot (highlighted by the arrow) indicates the observed location of the companion.}
\end{figure}

Figure \ref{fig:combined} shows the simulated brightness of the dust produced by SMACK during after 5 Myr, scaled for comparison with Figure 1 of \citet{Lagrange2016} and Figure 3 of \citet{Kalas2015}. The SMACK-produced dust population exhibits a brightness asymmetry at pericenter (Figure \ref{fig:combined}a) as well as a faint extension on the apocenter side (Figure \ref{fig:combined}b), although it does not reproduce the diffuse emission on the eastern side of the disk suggested by \citet{Kalas2015}. 

The scattered-light brightness enhancement at pericenter is a signature of an eccentric ring \citep{Wyatt1999a,Pan2016}, indicating that the orbits of the disk particles in the simulation, initially assigned eccentricities $<0.01$ and random longitudes of node and arguments of pericenter, have become more eccentric and apsidally aligned due to their secular resonance with the distant companion \citep[][see Section \ref{sec:time}]{Li2014}. The extended ``tail'' seen on the apocenter side of the disk is also a signature of an eccentric ring, in which high-eccentricity dust grains are produced near the ring's pericenter on apsidally aligned orbits, and then observed as they travel to and from their distant apocenters \citep{Lee2016}. 
 
The relatively low inclination of the companion relative to the disk results in a flat, narrow ring, with no appreciable vertical extension after 5 Myr, and the observed relative inclination between the companion and the disk ($\sim21^{\circ}$) is reproduced by our choice of $5^{\circ}$ relative inclination and the $85^{\circ}$ viewing inclination. The observed exterior companion is therefore able to reproduce four of the observed morphological features of the disk with no requirement for a second companion. 

The physical mechanisms described above allow us to place constraints on the orbit of the companion. The relative position angle of the companion and the disk's line of nodes on the sky is related to the inclination of the companion's orbit relative to the plane of the disk, but this relationship in complex and also depends on the longitude of nodes and argument of pericenter of the companion relative to the disk. In addition, these angles change with time, as the gravitational perturbations from the companion cause the orbits of the disk particles to precess coherently together (see Section \ref{sec:time}). Instead, we can use the disk's morphology to place constraints on the companion's orbit. For example, the pericenter of the companion cannot be too close to the outer edge of the disk or the companion's chaotic zone will truncate the disk. As a rough estimate, we can calculate the relationship between the radius of the inner edge of the companion's chaotic zone, $r_{\rm z}$, and the companion's pericenter distance, $r_{\rm pl}$, using the analytically derived classical chaotic zone relationship for circular orbits \citep{Wisdom1980}, $(r_{\rm pl}-r_{\rm z})/r_{\rm pl} = 1.3\mu^{2/7}$, where $\mu$ is the companion-to-star mass ratio. If we set the inner edge of the companion's chaotic zone to be the outer edge of the disk, $r_{\rm z}=100$ au, the minimum pericenter location for the companion is $r_{\rm pl}\approx138$ au. This constraint contains a degeneracy between the companion's eccentricity and semi-major axis, $a_{\rm pl} (1-e_{\rm pl})\gtrsim138$ au. 

There also exists an upper limit on the companion's semi-major axis, as the secular timescale must be less than the age of the system for the companion's secular perturbations to produce the observed asymmetries in the disk. If we constrain the secular timescale to be at least 10 Myr, the companion's semi-major axis and eccentricity are constrained by $a_{\rm pl} \sqrt{1-e_{\rm pl}^2}\lesssim661$ au (see Section \ref{sec:time}). 

\begin{figure}
\includegraphics[width=\linewidth]{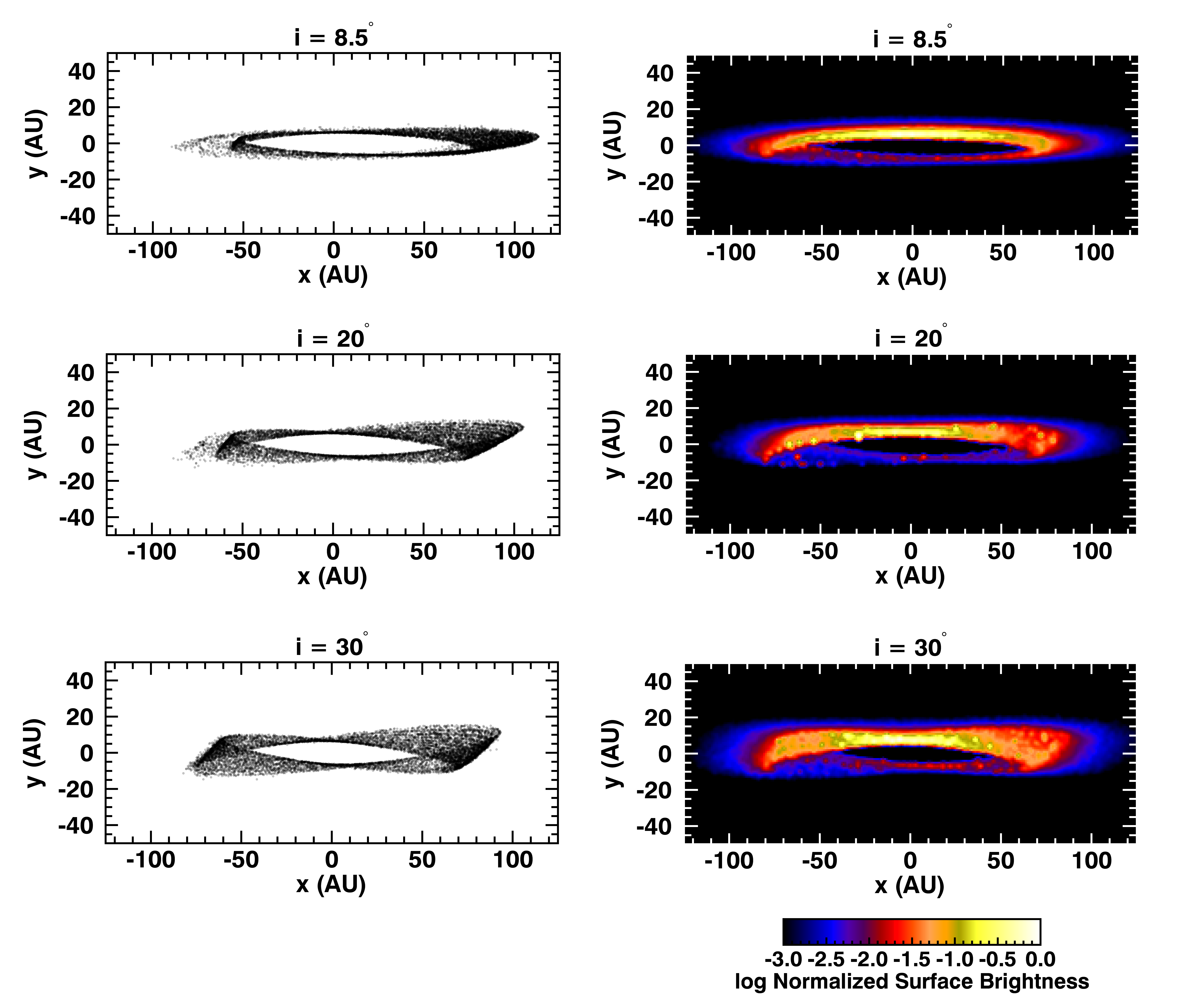}
\caption{\label{fig:verticalextent} Left: Locations of the parent bodies in each disk at 10 Myr. Each disk is inclined $5^{\circ}$ from edge-on and perturbed by a companion with inclination $8.5^{\circ}$, $20^{\circ}$, or $30^{\circ}$.  Right: Corresponding simulated surface brightness maps of each disk. The vertical extent of the disk increases with companion inclination, as well as with time.}
\end{figure}

We can use the disk's vertical extent to place an upper limit on the mutual inclination between the plane of the disk and the companion's orbit. We ran three collisionless $N$-body simulations of the disk perturbed by a 11 $\rm M_{\rm Jup}$ companion with the same orbital parameters as in the SMACK simulation, but varying the mutual inclination between the companion and disk to $8.5^{\circ}$, $20^{\circ}$, or $30^{\circ}$. The $N$-body particles represented the parent bodies in the disk. We simulated the production of dust grains by generating and recording one dust orbit matching the location and velocity of each parent body every 10 years during each simulation. We then generated 100 dust grains from each dust orbit and produced simulated brightness images using the procedure described in Section \ref{sec:dust}. Figure \ref{fig:verticalextent} shows the simulated brightness of each disk, viewed $5^{\circ}$ from edge-on, at 10 Myr, as well as the location of the parent bodies in each disk. Perturbations from a higher companion inclination produce a larger vertical extent in the disk. Resolved images of the system show a flat disk, indicating that the companion's inclination relative to the disk must be $i\lesssim20^{\circ}$. Our SMACK simulation demonstrates that the observed exterior companion can excite the necessary eccentricities in the ring within the age of the system without creating a significant vertical extension if the companion has a moderately large eccentricity ($e\approx0.7$) but a small inclination ($i\approx8.5^{\circ}$). Given these orbital parameters, the companion's semi-major axis would need to be $\sim700$ AU to match its observed projected position.

\section{Constraining the Companion's Orbit}
\label{sec:time}

\begin{figure}
\includegraphics[width=\columnwidth]{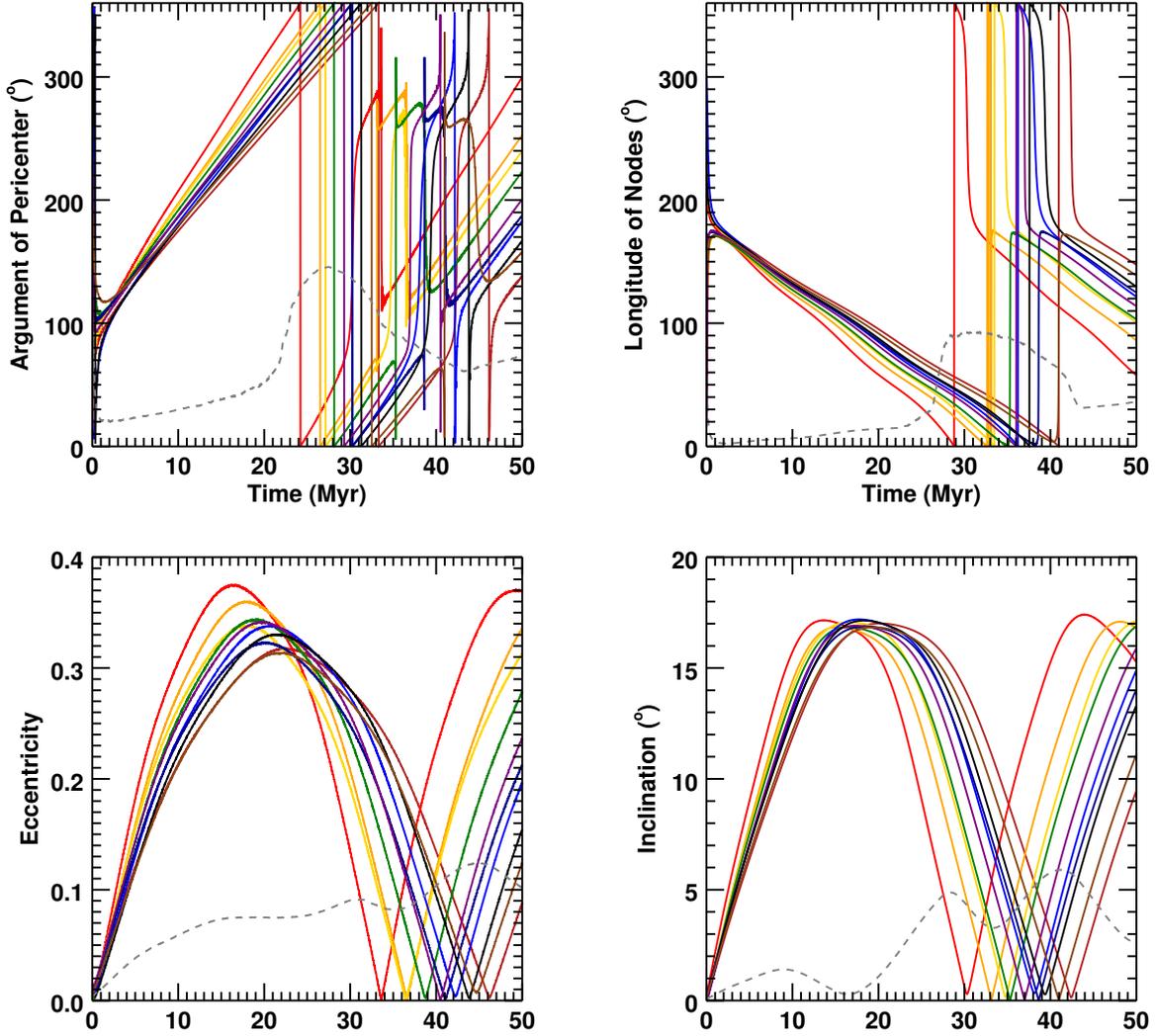}
\caption{\label{fig:alignment} Time evolution of the argument of pericenter, longitude of nodes, eccentricity, and inclination of ten randomly chosen particles in a collisionless $N$-body simulation over 50 Myr. The dashed grey line in each plot indicates the standard deviation of the given orbital element for all the particles in the disk. The rough convergence of the argument of pericenter and longitude of nodes produces a coherent ring of particles. The eccentricities and inclinations of the particles oscillate with time.}
\end{figure}

At time $t=0$ yr in our SMACK simulation of the HD 106906 system, the disk particles have eccentricity $\leq0.01$ and longitudes of node and arguments of pericenter distributed randomly between $0-360^{\circ}$, forming a circular belt. By time $t=5$ Myr, gravitational perturbations from the companion have increased the average eccentricity of the particles to $\sim0.18$ (Fig. \ref{fig:alignment}). Increasing the average eccentricity of the particles alone would only produce a broader circular disk, but the secular resonance formed with the companion also cause the particles' longitudes of node and arguments of pericenter to converge. In other words, the orbits of the disk particles begin to apsidally align, producing a coherent eccentric ring. Fig. \ref{fig:alignment} illustrates this with plots of the time evolution over 50 Myr of the longitudes of node, arguments of pericenter, eccentricities, and inclinations of ten randomly chosen disk particles in a collisionless $N$-body simulation with the same system parameters as the SMACK simulation described in Table \ref{tab:init}. We used 10,000 particles to represent the disk, simulated with the Wisdom-Holman integrator in REBOUND with collisions turned off. The orbits of these ten randomly chosen particles become roughly apsidally aligned within $\sim4$ Myr. This behavior is consistent with the hierarchical nearly coplanar secular evolution, investigated in \citet{Li2014}, which showed that the resonance angle for low-inclination companions is the sum of the longitude of nodes ($\Omega$) and the argument of pericenter ($\omega$). The test particles, although they were initially assigned random values of $\Omega$ and $\omega$, are captured into resonance with the companion, which both pumps up their eccentricity and aligns their orbits, forming an eccentric disk.

The secular precession timescale of particle in the disk due an external planetary-mass companion with mass $M_{\rm pl}$ and eccentricity $e_{\rm pl}$ is defined as \citep{Naoz2016}
\begin{equation} \label{eq:tsec} t_{\rm sec}\sim \frac{16}{30\pi} \frac{M_{*}+M_{\rm pl}}{M_{\rm pl}}\frac{P_{\rm pl}^2}{P_{\rm d}}(1-e_{\rm pl}^2)^{3/2}, \end{equation} 
where $P_{\rm pl}$ is the companion's period around the star (mass $M_{*}$) and $P_{\rm d}$ is the period of particles in the disk. The eccentric companion induces gravitational perturbations which result in the disk particles orbiting in a resonance, where the disk particles' longitude of pericenter $\varpi=\omega+\Omega$ is the resonant angle \citep{Li2014}. 

The precession timescale can be used to place an upper limit on the companion's semi-major axis. In order to perturb the entire disk (down to its inner edge at $\sim 50$ au) within the age of the system $t_{\rm age}\approx10$ Myr, the inner edge of the disk must have experienced at least one half-cycle of secular perturbation, so the secular timescale at 50 au must be $\frac{1}{2} t_{\rm sec}\lesssim t_{\rm age}$. Using a stellar mass of 2.5 ${\rm M}_{\odot}$ and a companion mass of 11 ${\rm M_{Jup}}$, and Equation \ref{eq:tsec}, this yields
\begin{equation} a_{\rm pl}\sqrt{1-e_{\rm pl}^2}\lesssim661~{\rm au}. \end{equation}

\section{Summary and Conclusions}
\label{sec:conclusions}

We have shown that the observed exterior companion in the HD 106906 system can shape the system's debris disk into a flat, eccentric, dust-producing ring and reproduce its observed morphological features and asymmetries. Our SMACK simulations also allow us to place constraints on the orbit of the companion using the morphology of the disk.

While we have demonstrated that the perturbations from the observed, exterior companion can excite eccentricities in the HD 106906 ring, alternative mechanisms for eccentricity excitation also exist. For example, a second companion on an eccentric orbit interior to the debris ring could force an eccentricity on the ring. Future simulations investigating the plausibility of this scenario may be able to constrain the orbit of the outer companion based on stability requirements. 

Constraining the orbit of HD 106906b could have implications for its formation scenario. Prior to the publication of resolved images of the disk, it was suggested (using $N$-body simulations) that the companion formed interior to the disk and was scattered onto a highly eccentric orbit \citep{Jilkova2015}. This study concluded that the disk can survive perturbations by a companion with an apocenter distance of 650 au and a pericenter distance interior to the disk if the companion's inclination is $\gtrsim10^{\circ}$. However, this resulted in a significantly vertically perturbed disk by 10 Myr, regardless of the companion's inclination. Our simulations indicate that a companion with an orbit completely exterior to the disk can reproduce the observed asymmetries without vertically extending the disk, supporting the scenario in which the companion formed \textit{in situ}. 

A more thorough exploration of the parameter space may be able to place further constraints on the companion's orbit using the observed geometry of the system. The methodology we presented in this work can also be generalized to other debris disk observations to explore whether their observed asymmetries could be explained by the presence of an undetected distant exterior companion, and investigate how the morphology of these disks could constrain the orbit of their exterior perturbers. Other debris disk systems with exterior massive perturbers are likely not uncommon; surveys indicate that $\sim25\%$ of debris disks exist in binary or triple star systems \citep{Rodriguez2012}. Planetary-mass exterior companions like HD 106906b may be responsible for the asymmetries in observed debris disks such as HD 61005 \citep{Hines2007,Esposito2016} and HD 15115 \citep{Rodigas2012,Schneider2014, MacGregor2015a}, for example.

\vspace{1em}

Numerical simulations were performed on the Memex High Performance Computing Cluster at the Carnegie Institution for Science. Erika Nesvold was supported by the Carnegie DTM Postdoctoral Fellowship. Smadar Naoz acknowledges partial support from a Sloan Foundation Fellowship. The authors wish to thank the anonymous referee for a prompt and helpful review.

\bibliographystyle{apj}

\end{document}